# Directly measuring the spatiotemporal electric field of ultrashort Bessel-X pulses

P. Bowlan[1*], R. Trebino[1], H. Valtna-Lukner[2], M. Lõhmus[2], P. Piksarv[2], and P. Saari[2]

[1] *Georgia Institute of Technology, School of Physics 837 State St NW, Atlanta, GA 30332 USA*
[2] *University of Tartu, Institute of Physics, 142 Riia St, Tartu, 51014 Estonia*
[*] *Corresponding author's e-mail address: PamBowlan@gatech.edu*

**Abstract:** Using SEA TADPOLE with µm-range spatial and fs-range temporal resolution, we report the first direct spatiotemporal measurements of ultrashort Bessel-X pulses. We demonstrate their propagation invariance and superluminal velocity and verify our results with simulations.

## 1. Introduction

Bessel-X pulses are of great interest because they propagate in vacuum or linear media over large distances like an optical bullet – without exhibiting any diffraction or spread. Bessel pulses have many applications, such as filament or plasma generation and have even been used for cell transfection. Bessel-X pulses, a type of axially symmetrical localized wave (see [1-4] and references therein), are a wideband wave-packet of coaxial zeroth-order Bessel beams having a proportionality relation between their temporal frequency and the axial wave-number of the individual Bessel beams. Their three dimensional intensity profile $I(x,y,z,t)$ comprises a strong central spot surrounded by weaker interference rings whose diameter increases with distance away from $t = 0$, so they look like two cones extending out in time from the origin. An $x$-$t$ or $x$-$z$ slice $I(x,t)$ or $I(x,z)$ of the pulse resembles the letter "X". The field (and also the intensity) of the Bessel-X pulse propagates in the axial ($z$) direction rigidly, and the propagation speed of the pulse is superluminal (in vacuum greater than c) which is not in violation of Einstein's causality [4]. It is important to measure these pulses, not only to observe their interesting and useful properties, but also to aid in their generation and application. But Bessel-X pulses have a complex spatiotemporal shape, so their complete and accurate measurement requires a sophisticated spatiotemporal technique with high temporal and spatial resolution.

Experimentally the spread-free propagation of the central spot of a Bessel-X pulse, generated by a special holographic element, was first studied in [3]. The X-like profile and the superluminal propagation was for the first time demonstrated in [4], in which an annular slit and a lens were used to generate corresponding correlations using a white light source. The authors of [5] studied propagation of an ionization front in argon gas caused by the central spot of an intense Bessel-X pulse generated from a 70 fs laser pulse by a conical lens (axicon). This paper reported the pulse's group velocity $v_g = 1.1c$. And, to our knowledge, no one has ever made a direct spatiotemporal measurement of the electric field of a Bessel-X pulse.

In this paper we report the first, direct measurements of "snapshots in flight," or spatiotemporal slices of the X-like profile of a femtosecond Bessel-X pulse. Our results show propagation invariance over ~ 7 cm as well as the superluminal velocity of the Bessel-X pulse. To make these measurements we used the linear-optical interferometric technique SEA TADPOLE [6,7], and we verified them with numerical simulations.

## 2. Experimental results and numerical simulations

To generate the Bessel-X pulse we used a fused silica axicon with an apex angle of 176 degrees and a KM Labs Ti:Sa oscillator with 40nm of bandwidth (FWHM). The spot size of the beam at the axicon was 4mm (FWHM). A detailed description of our measurement device SEA TADPOLE can be found in [6-7]. Briefly, this device involves sampling a small spatial region of the Bessel pulse with a single-mode optical fiber (having a mode diameter of 5.4µm, which is therefore our spatial resolution) and then interfering this with a reference pulse in a spectrometer to reconstruct $E(\lambda)$ for that spatial point. Then to find the spatial dependence of the field we scan the fiber axially (in $x$) throughout the cross section of the Bessel-X pulse, so that $E(\lambda)$ is measured at each $x$, yielding $E(\lambda,x)$. This field can be Fourier transformed to the time domain to give us $E(t,x)$. In order to measure the z (propagation direction) dependence of the spatiotemporal field, the axicon can be translated along the propagation direction to bring it nearer or further from the sampling point (the fiber). Note that we could also scan the fiber along the $y$ dimension to measure $E(\lambda,x,y)$, but our Bessel-X pulses were approximately symmetric about the $z$-axis, so this was not necessary. Our temporal resolution was 8.9fs, and we used zero-filling to decrease the point spacing to 4.6fs. As explained in [6], SEA TADPOLE measures the spectral phase difference between the unknown and the reference pulse and for these measurements we placed extra glass in the reference arm to cancel out the group delay dispersion (GDD) introduced by the center thickness of the material in the axicon. Therefore our measurements reflect the spatiotemporal phase introduced by the axicon.

We measured $E(\lambda,x)$ at several different values of z or axial distances from the front surface of the axicon by translating it along the beam's propagation direction, and three of these measurements are shown in Fig.1.

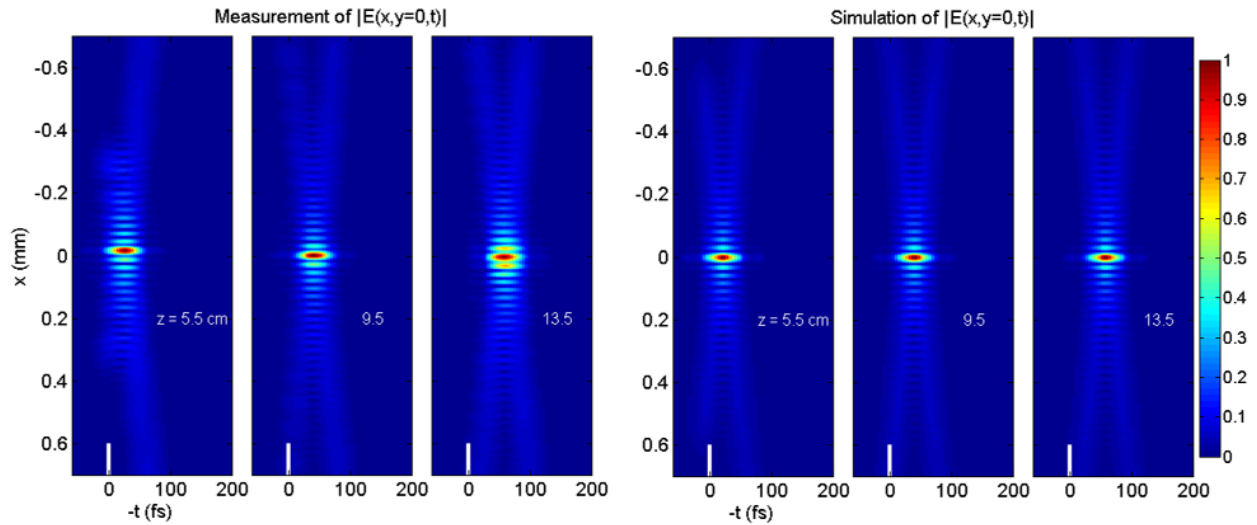

Figure 1. Left: The measured field amplitude at three different distances (*z*) after the axicon. Right: and the corresponding simulations. Color is intensity as indicated by the color bar and we normalized each field to have a maximum of 1.

We also performed numerical simulations, and these results are shown on the right in Fig.1. The two are in good agreement except that the wings in the z=5.5cm image are shorter in the measurement. This is because axicons are difficult to machine perfectly and therefore the tip of the cones are always distorted and though we have tried to account for this in our simulations, it is difficult to model perfectly. Our device also measures the spatiotemporal phase, but because we compensated for the glass in the axicon, this phase is transform limited, or the color of the pulse is the same everywhere in *x* and *t*. So here we only show the amplitude.

There are several interesting features in this data. The central maximum of the pulse has a width of ~ 20 μm, which remains essentially unchanged in shape from $z = 5$cm through $z = 13.5$cm, and in our measurement at $z = 13.5$, the interference pattern is just beginning to change. The discrepancy in size of the wings (left arms of the "X") and the slight difference in the central intensity pattern at z=13.5cm are due to defects in the axicon's surface.

Also, the Bessel X-pulse's superluminal speed is apparent in these plots. SEA TADPOLE measures the pulse's arrival time with respect to the reference pulse which is just a simple Gaussian in this case that travels at the speed of light (*c*). Therefore if the Bessel X-pulse were traveling at the speed of light then at each *z* its spatiotemporal intensity would be centered at the same time (here $t = 0$ and emphasized with the white line), but it is easy to see that this is not the case. From our axicon's apex angle (and from the simulations), we find that the Bessel X-pulse's speed should be $1.00013c$. Therefore, over a distance of 8cm, the Bessel X-pulse would lead our moving reference frame (the reference pulse) by 35 fs. In our results, the center of the pulse is time-shifted by 32fs forward between $z = 5.5$cm and $z = 13.5$cm, which is in good agreement with our theoretical prediction. To verify this result, we repeated this experiment five times and consistently measured the time shifts predicted for this axicon. We also realigned the axicon in between these trials to assure that this delay was not due to (or significantly affected by) misalignment of the axicon's scanning stage.

In conclusion, using SEA TADPOLE, we have made the first (to our knowledge) direct spatiotemporal measurements of Bessel-X pulses, and we verified these results with simulations. We demonstrated both the propagation invariance of the Bessel-X pulse and its superluminal axial group velocity which we determined to be around 1.00012c – within 0.001% error of the expected value.

Note added before uploading to arXiv: according to a private communication from P. Di Trapani, a study with a similar purpose but using a different method was published on April 6, 2009 by F. Bonaretti *et al*, arXiv:0904.0952 and presented at the 2nd International Workshop on Filamentation, Paris September 22-25, 2008.